\documentclass{aa}
\usepackage{graphics}
\newcommand{\bly}{$\beta$ Lyrae\,}  

\begin{document}

   \thesaurus{08         
              (08.02.4;  
               08.03.2;  
               08.09.2  
			)
                }

\title{Resolving the radio nebula around \bly} 

\author{G. Umana\inst{1} \and 
          P.F.L. Maxted\inst{2} \and
          C. Trigilio\inst{1}  \and
         R.P. Fender\inst{3,4} \and
           F. Leone\inst{5}  \and
         S.K. Yerli\inst{3}  
} 

\offprints{G. Umana}

\institute{
Istituto di Radioastronomia del C.N.R., Stazione VLBI di Noto, 
C.P. 141,  Noto (SR), Italy  
\and
University of Southampton, Department of Physics and Astronomy, Highfield, 
   Southampton, SO17 1BJ, UK  
\and
 Astronomy Centre, Department of Physical Sciences, University of  
   Sussex, Falmer, Brighton BN1 9QH, UK    
\and
Astronomical Institute `Anton Pannekoek',Center for High-Energy Astrophysics, University of Amsterdam,  Kruislaan 403, 1098 SJ Amsterdam,  The Netherlands
\and 
Osservatorio Astrofisico di Catania, Viale A. Doria 6, 
 95125 Catania, Italy
}

\date{Received To be inserted later, accepted To be inserted later}   
  
\maketitle  
  
\begin{abstract} 
In this paper we present   high spatial resolution radio images of the
puzzling binary system \bly{} obtained with MERLIN at 5\,GHz. We find a nebula
surrounding the binary with a brightness temperature of $(11\,000\pm700)$\,K
approximately 40~AU across. 
This definitively confirms the thermal origin of the radio emission, which is 
consistent with emission from the wind of the B6-8\,II component (mass loss of 
order of $ 10^{-7}~M_{\sun} \mathrm{yr}^{-1}$), ionized by the radiation field 
of the hotter companion.
This nebula, surrounding the binary, is the proof that \bly evolved in a 
non-conservative way, i. e. not all the mass lost by the primary is accretted 
by the secondary, and present measurements indicate that  almost 0.015 
$M_{\sun}$ had been lost from the system since the onset of the 
Roche lobe overflow phase.

Moreover, the nebula is aligned with the jet-like structures inferred from 
recent optical measurements, indicating a possible connection among them. 
   \end{abstract}  
  
\section{Introduction}  
  
\bly ($3.4 \leq m_{V} \leq 4.4$, $P=12.91^\mathrm{d}$) is a non-degenerate,
semi-detached interacting binary system, located at a distance of 270$\pm 38$\,pc (Perryman et al.\cite{perryman}). The primary component is a B6-8\,II star in
contact with its Roche lobe resulting in rapid mass transfer ($5\times 10^{-5}
M_{\sun}\mathrm{yr}^{-1}$) to its massive, unseen companion (Harmanec \cite{Harmanec90}).  The
nature of the secondary has been the subject of much speculation,  from a
flattened B5 star to a black hole with accretion disk.  A B0V star embedded in a
geometrically and optically thick accretion disk or a circumstellar shell, is
currently the more favoured  hypothesis (Hubeny $\&$ Plavec \cite{Hubeny}, Harmanec \cite{Harmanec92}). The presence
of circumstellar plasma surrounding both components  is indicated by stationary
optical and UV emission lines (Batten \& Sahade \cite{Batten}; Hack et al. \cite{Hack}) as
well as from the analysis of UV light curves (Kondo et al. \cite{Kondo}). 
 Although \bly has been observed extensively at other wavelengths for
decades, radio observations are still too scarce to form a clear picture of its
radio properties. The source was firstly detected by Wade \& Hjellming (\cite{wade}) at
2.7 and 8.1\,GHz.  During a successive monitoring program (Gibson 1975) the
source always exhibited a spectrum consistent with thermal radio source with a
radius $\sim 50$~AU, if a temperature of typical of an \ion{H}{II} region 
($10^{4}$\,K)
is assumed. However, Wright \& Barlow (\cite{Wright}) have noted that the observed slope
of the radio spectrum ($\alpha$=0.96) is  intermediate between that expected for
a simple \ion{H}{II} region  and a stellar wind, implying that the physics underlying
the  radio emission of \bly is more complicated than assumed in
either of these models. 

 In an attempt to distinguish the various components of \bly which
may be contributing to the radio emission, we obtained high-spatial resolution
radio maps of \bly using MERLIN. In this paper we describe the
results of these observations and briefly discuss their implications.

\begin{figure} 
\resizebox{\hsize}{!}{\includegraphics{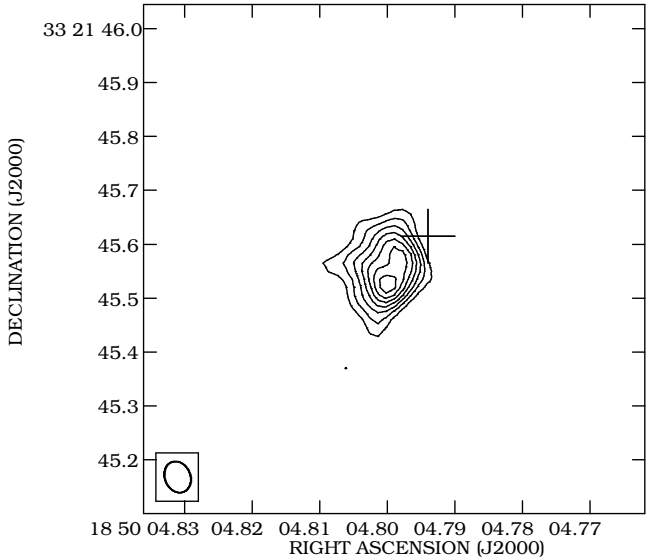}} 
      \caption{The 6\,cm radio map  of \bly. The contour plot shows the
results of three separate 18 hour observations of \bly combined into
a single image. Levels of 0.18 (3$\sigma$), 0.24, 0.30, 0.36, 0.42, 0.48 and 0.54 mJy are shown. The position measured by Hipparcos is indicated with a large
cross. The small inset ellipse is the  half-power contour of the beam ($60
\times 47 $mas). }
  \label{}
\end{figure}

\section[]{Observations and Results}  
 
 We observed \bly at 6\,cm (4.994\,GHz) using the Multi-Element Radio
Linked Interferometer Network (MERLIN)\footnote{ MERLIN is a national facility
operated by the University of Manchester at the Nuffield Radio Astronomy Laboratories, Jodrell Bank, on behalf of the Particle Physics and
Astronomy Research Council (PPARC)} array on 1996  December 8, 9 and 16,
from 04:00 to 22:00 UT. 

 The observations were performed in phase referencing mode, which is the
standard MERLIN observing mode for weak sources. This technique allows a
better  calibration of phase variation by observations of a bright phase
calibrator interleaved with the observations of the target source. The phase
calibrator 1846+322 was used and the flux density scale was  determined by
observing 3C286, whose flux at 4.994\,GHz is assumed to be 7.086\,Jy.

\begin{table} 
\caption{Parameters of the radio nebula around \bly.}
 \begin{tabular}{@{}lr@{}}
Parameter & \multicolumn{1}{c}{Value} \\ 
\hline
Flux density			& $3.2 \pm 0.2			$ mJy \\
Size~(FWHM)			& $(145 \pm 12)\times(100 \pm 8)$ mas \\
Orientation~(major axis)	& $156.5\degr \pm 4\degr $ \\
\noalign{\smallskip}
 Radio position (J2000)  \\
RA           & $18^\mathrm{h} 50^\mathrm{m} 04\fs 800 \pm 0\fs 0060$\\
Dec          & $+33\degr      21\arcmin     45\farcs 554 \pm 0\farcs 004$\\
Epoch        & 1996.9  \\
\noalign{\smallskip} 
Optical position$^{a}$ (J2000) \\
RA           & $18^\mathrm{h} 50^\mathrm{m} 04\fs 794 \pm 0\fs 0002$ \\
Dec          & $+33\degr      21\arcmin     45\farcs 615 \pm 0\farcs 003$\\
\hline
\multicolumn{2}{l}{$^{a}$ Hipparcos position reduced to same epoch} \\
\multicolumn{2}{l}{~~including proper motion.}\\
\end{tabular}
\end{table}

 Data were edited and amplitude calibrated using the OLAF software package.
The phase calibration, mapping process and successive analysis were performed
using the NRAO, {\bf A}stronomical {\bf I}mage {\bf P}rocessing {\bf S}ystem.
To achieve the highest possible signal-to-noise ratio the mapping process was
performed using natural weighting and the  CLEAN algorithm was used to extract
information as close as possible to the theoretical noise limit. 

 We detected the source at all three epochs, corresponding to three different
orbital phases. All three maps revealed a quite extended source, elongated
slightly in the N-S direction.   The noise level in the maps measured from 
several areas of blank sky is $6 \times 10^{-5}$Jy/beam, which is consistent
with the expected theoretical noise. There are no significant differences
between the maps observed at each epoch and so we combined the data from the
three observing runs to form the radio map shown in Fig.~1. 

 The source morphology is dominated by a compact structure,  aligned 
approximately N-S, probably
 ($< 3\sigma$ level), embedded in a more diffuse 
nebula. 
The profile of the radio nebula along its
principal axes is shown in Fig.~2. From the figure is evident that 
most of the 6~cm emission  comes from the central region, with negligible
contribution from the more diffuse nebula,  thus 
proving that a two dimensional Gaussian fit (dotted line) to the central part 
gives a fair description of the radio source.

The position, flux density and 
angular size of the source are thus derived by fitting the two dimensional Gaussian
brightness distribution to the map. The values and their
uncertainties, derived following Fomalont (\cite{Fomalont}), are given in Table~1. The
optical position of \bly reported by Hipparcos, allowing for the
proper motion, is also given.  
Differences between radio and optical positions ($\Delta \alpha=0.075\arcsec, 
\Delta \delta=-0.061\arcsec$) may be probably related to the overall position error to be associated to the source used as phase calibrator,
reported to be of the order pf $0.1\arcsec$ (Wilkinson et al. 1998).

 MERLIN is sensitive to structures up to 3.5\arcsec\ at 6\,cm. No structure is
seen outside of the nebula up to these scales, which would suggest that all of
the flux from the nebula is detected in these maps. The difference between the
flux measured here and previous measurements (e.g. Mutel et al. \cite{Mutel}; Leone, Trigilio \& Umana \cite{Leone}) is not surprising as \bly is known to show variability at
radio wavelengths.

\begin{figure} 
\resizebox{\hsize}{!}{\includegraphics{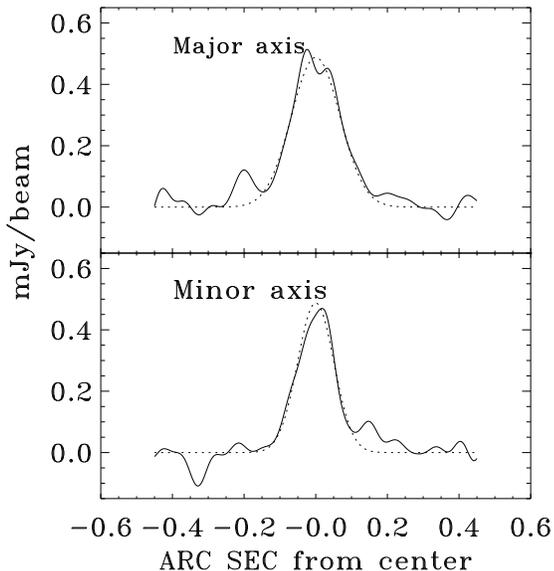}} 
      \caption{Cross sections of the radio nebula along its principal axes (solid
lines) together with the best fitting two dimensional Gaussian intensity
distribution (dotted line).}
  \label{borra}
\end{figure}

\section[]{Discussion and Conclusions} 
 The radio observations here  discussed are direct confirmation of a large plasma cloud
surrounding \bly due to mass loss from the system. The angular size of
the nebula corresponds to a linear  size of approximately 40~AU, much
larger than the binary system itself ($\sim 0.5$~AU). As the
source has been resolved we can derive a brightness temperature $T_\mathrm{B}$ 
as follows:
\begin{equation} 
T_\mathrm{B}=\frac{1.36 \times 10^{6} S \lambda^{2}} {\theta_{1} \theta_{2}}
~\mathrm{K}=11\,000\pm700 ~\mathrm{K}
\end{equation} 
where $S$ is the flux in mJy at the 5\,GHz,  $\lambda$ is the observing
wavelength in cm and $\theta_{1}$ and $\theta_{2}$ are the angular size of
source in milliarcseconds. The derived  brightness temperature  implies  a thermal origin for the radio emission.

It can be argued that the source has, in reality, a much more complex morphology
that the present MERLIN observations can not resolve. If, for example, most of 
the flux is coming from two small components, with thus a much higher brightness
temperature, the simple one-component model, we are assuming, will be no 
suitable.
Still, in the former case, a higher brightness temperature would imply a non-thermal mechanism operating in the system. The presence of non thermal
component has been inferred in about 30$\%$ of the early type radio emitting 
stellar systems (Bieging et al. 1989). All of these stars always exhibited spectral indices
$\alpha \leq 0$ ($S_{\nu} \propto \nu^{\alpha}$),
while past (Gibson, 1975) and more recent (Umana et al. 2000) radio observations of \bly{}
point out a thermal spectrum ($\alpha > 0$)
 for  the system.

Jameson \& King (\cite{Jameson}) modelled the radio source associated to \bly 
as a \ion{H}{II} region. However, there is ample evidence, from both optical 
(Etzel \& Meyer, 1983)  and ultraviolet spectroscopy (Mazzali et al. 1992), 
of the presence of stellar wind in one or both components of the 
system, which can be responsible of the observed radio 
emission.

Mazzali et al. (\cite{Mazzali}) have developed a two-wind model  to explain the peculiar UV spectrum of \bly{}, consisting of "superionized" resonance lines, typical of hot-star winds, plus lines of moderately ionized species.
The observed ultraviolet emission features can be ~modelled allowing the coexistence of two winds with different dynamical properties: 
a fast, but tenuous wind, associated to the hotter secondary component 
($\dot {M_\mathrm{h}}=5 \times 10^{-8} M_{\sun}\mathrm{yr}^{-1}$, 
$v_\mathrm{h}=1470 ~\mathrm{km~s}^{-1}$)  and a denser, but slower wind, 
associated to the cooler B6-8\,II primary 
($\dot {M_\mathrm{c}}= 7.7 \times 10^{-7} M_{\sun}\mathrm{yr}^{-1}$, 
$v_\mathrm{c}= 390~\mathrm{km~s}^{-1}$). 
Standard formulas for thermal radio emission from an expanding wind have been
derived by  Panagia \& Felli ({\cite{Panagia75}) and
Wright \& Barlow  (\cite{Wright}).  They showed that the observed radio flux ($S$) is related to the dynamical parameters of the wind through the relation:
\begin{equation}
S \propto \left(\frac{\dot{M}}{v_{\infty}}\right)^{4/3}
\label{flux}
\end{equation}
where $\dot{M}$ and $v_{\infty}$ represent the mass-loss rate and the terminal
velocity of the wind.

We may ask under which conditions the thermal radio emission from the winds
of the two early-type components of a binary system has the same radio 
properties as a single symmetric wind and thus the relation (\ref{flux}) 
can be applied.

 The effects of binarity on thermal radio emission from early-type systems
have been recently studied by Stevens (1995). Effects  of possible 
interactions between winds, such as extra source of emission due to the shocked
gas, are mostly function of the 
momentum ratio of the two winds and they would  be negligible in systems,
where one component has a dominant wind.
The two-wind model outlined by Mazzali et al. (1992) foresees a case of a system when 
only one wind is dominant. Moreover, since the winds of the two stars will contribute to the measured flux according to the ratio of the relative factors  
$(\frac{\dot{M}}{v_{\infty}})$, allowing for difference in mass-loss and wind terminal velocity, this indicates that less than 0.5 \% of the observed flux is attributable to the BOV star's wind.

There is a class of stellar objects, the symbiotics, whose radio emission is interpreted in terms of a binary model, and where one of the two components has a dominant wind. 
Taylor \& Seaquist (\cite{Taylor84}) attributed the  observed radio emission 
to the stellar wind of the cool component of the system, which is ionized by 
the ultraviolet flux of the hot companion. 
They also showed that the geometry of the ionized radio emitting region of the 
intersystem material is determined by a single parameter,
which is function of the physical parameters of the binary: 
separation ($a$), Lyman continuum luminosity of the hot component ($L_{uv}$),  mass loss rate ($\dot {M}$) and  velocity ($v$)
of the dominant wind.
If the wind  is completely ionized, the ionized region shows the same radio properties as a spherically symmetric ionized wind.
In the following we  will check if this scenario is suitable also for \bly. \\
By applying equation 14 from Taylor \& Seaquist (\cite{Taylor84}) and assuming
 a binary separation of 61 solar radii (Harmanec, 1990), 
we obtain that the wind of the B6-8\,II is completely
ionized if 
\begin{equation}
  L_{uv}\geq  5.8 \times 10^{62}  (\frac{\dot {M}}{v})^{-2} 
\sim 1.8 \times 10^{45}~~~\mathrm{photons~s}^{-1} 
\end{equation}
where the values of $ \dot {M}$ and $v$ for the B6-8\,II as determined by Mazzali et al. (1992) have been adopted.
This requirement is well satisfied by the flux of Lyman continuum photons of 
the bright secondary component B0V, 
$L_{uv}=4.26 \times 10^{47}~\mathrm{photons~s}^{-1}$ 
(Panagia \cite{Panagia73}), on the contrary of the B6-8\,II, 
and we can safely apply relation (\ref{flux}).

This formula has been derived for the case of spherically symmetric winds.
However, Schmid-Burgk (\cite{Burgk}) showed that the mass loss rate derived 
from radio flux densities is quite insensitive to source geometry and only in 
very extreme cases of deviation from spherical symmetry a geometry-dependent
correction factor must be applied. Thus even if the source morphology appears 
to be slightly elongated (axis ratio 1.45), we can use  them quite confidently.

We, therefore, can estimate the mass loss by using the relation:
 \begin{equation}
\dot {M}= 6.7 \times 10^{-4} v_\infty 
S_\mathrm{6cm}^{3/4} D_\mathrm{kpc}^{3/2} (\nu\times g_\mathrm{ff})^{-0.5}  
~~~~ M_{\sun}\mathrm{yr}^{-1}
\end{equation}
where full ionization and cosmic abundances have been assumed,  $v_{\infty}$ is the terminal velocity of the wind, $S_\mathrm{6cm}$ the observed flux density 
at 6~cm in mJy, $D_\mathrm{kpc}$  the distance of the system, in  kpc. 
$g_\mathrm{ff}$ represents 
the free-free  Gaunt factor that, following Leitherer \& Robert (1991), can be    approximated  with:
\begin{equation}
g_\mathrm{ff}=9.77(1+0.13 \log{\frac{T^{3/2}}{\nu}})
\end{equation}
where $T$, in Kelvin, is the wind temperature.\\

By assuming the  stellar wind velocity of  the B6-8\,II component ($v\sim 400~  \mathrm{km~s}^{-1}$, Mazzali et al. \cite{Mazzali}) and constant temperature equal to the brightness temperature, derived by the present observations 
($T=T_\mathrm{B}$), we
obtain a mass loss rate of 
 \begin{equation}
\dot {M}= 5.6 \times 10^{-7}~~~~ M_{\sun}\mathrm{yr}^{-1}.
\end{equation}

This result is close to the value obtained by Mazzali et al. (\cite{Mazzali}) 
for the B6-8\,II component by modelling UV spectral lines, i. e. 
$7.16 \times 10^{-7} ~M_{\sun}\mathrm{yr}^{-1}$ 
and indicates that the wind associated to the primary is predominant in the  
replenishment of the circumsystem material and constitutes the bulk of the 
system's mass loss.

\bly{} is approaching the end of a phase of strong  mass transfer between
components, which started 26000 years ago (De Greve \& Linnell \cite{Degreve}).
The presence of such a nebula is the proof that the system had evolved
in a non conservative way, i. e.,  with a  fraction of the material from the 
loser lost  by the system.  This was already foreseen by 
the evolutionary studies  of  De Greve \& Linnell (\cite{Degreve}) since 
their conservative solution lead to several discrepancies with the observed characteristics of the system. 

If we assume that during this phase of mass transfer
the mass-loss rate was constant ($5.6 \times 10^{-7}   M_{\sun}\mathrm{yr}^{-1}$) we estimate that almost 0.015 $M_{\sun}$ should have been lost by the system.

High resolution optical interferometry combined with extensive
photometry and spectroscopy has led  Harmanec et~al. (\cite{Harmanec96}) to conclude that
\bly contains jet-like structures perpendicular to the orbital plane
which are responsible for the bulk of the H$\alpha$ and \ion{He}{I}~6678 
emission.
This scenario, already proposed by Harmanec (\cite{Harmanec92}), is further supported
by recent spectropolarimetric observations of Hoffman et al. (\cite{Hoff}),
 who suggested the presence of a bipolar outflow
in \bly, perpendicular to the orbital plane (P.A. = $164\degr$) and probably associated to the accretion disk around the mass-gaining component.
     
The  radio image presented in this work gives a hint of structure within the nebula, even if the  obtained angular resolution did not allow us to trace the structure of the star's circumstellar matter at very small scale and thus to verify the association
 of the radio emitting features to the bipolar optical jets. One piece of information on this direction, provided by the present results, is that the radio  nebula is almost exactly aligned with these optical jet-like structures (P.A. = $156.5\degr \pm 4\degr$).

It may be more accurate to picture the outflow as a
stellar wind collimated by the thick accretion disk. The  appearance of the
circumbinary material would thus consist of a more extended component, due to the mass-loss from the system as a whole, plus an inner more compact component, related to the accretion disk created in the process of mass transfer from the B star to the hidden companion.  

To establish the feasibility of this scenario high resolution radio observations, at frequencies higher than 5~GHz,
aimed to probe the morphology of the inner structure of the radio nebula  are necessary. 
This observations, by allowing to look more inside the binary's outflow,  would provide further insight into the formation of the circumstellar matter and are of particular importance in understanding the connection  of the observed radio nebula
to the bipolar optical jets  reported by Harmanec et al. (\cite{Harmanec96})
 and  Hoffman et al. (\cite{Hoff}).

\acknowledgements{ 
We wish to thank the MERLIN staff for conducting the observations and, 
in particular, Dr. T. Muxlow for its help in the first stage of data 
reduction and some helpful suggestions.}

\label{lastpage}  
 \end{document}